\documentclass[aps,prb,twocolumn,superscriptaddress,showpacs]{revtex4}

\usepackage{graphicx}
\begin{document}

\title{Magnetic structure of antiferromagnetic NdRhIn$_5$}

\author{S. Chang}
\email{schang@lanl.gov}
\affiliation{Los Alamos National Laboratory, Los Alamos, NM 87545}
\affiliation{Physics Department, New Mexico State University, Las Cruces, 
NM 88003}

\author{P.G. Pagliuso}
\affiliation{Los Alamos National Laboratory, Los Alamos, NM 87545}

\author{W. Bao}
\affiliation{Los Alamos National Laboratory, Los Alamos, NM 87545}

\author{J.S. Gardner}
\affiliation{NRC Canada, NPMR, Chalk River Laboratories, Chalk River, Ontario, Canada K0J 1J0}

\author{I.P. Swainson}
\affiliation{NRC Canada, NPMR, Chalk River Laboratories, Chalk River, Ontario, 
Canada K0J 1J0}

\author{J.L. Sarrao}
\affiliation{Los Alamos National Laboratory, Los Alamos, NM 87545}

\author{H. Nakotte}
\affiliation{Physics Department, New Mexico State University, Las Cruces,
NM 88003}

\date{\today}

\begin{abstract}
The magnetic structure of antiferromagnetic NdRhIn$_5$ has been determined using neutron diffraction.  It has a commensurate antiferromagnetic structure with a magnetic wave vector $(\frac{1}{2}~0~\frac{1}{2})$ below $T_N = 11$~K. The staggered Nd moment at 1.6~K is $2.6~\mu_B$ aligned along the $c$~axis. We find the magnetic structure to be closely related to that of its cubic parent compound NdIn$_3$ below 4.6~K. The enhanced $T_N$ and the absence of additional transitions below $T_N$ for NdRhIn$_5$ are interpreted in terms of an improved matching of the crystalline-electric-field (CEF), magnetocrystalline, and exchange interaction anisotropies. In comparison, the role of these competing anisotropies on the magnetic properties of the structurally related compound CeRhIn$_5$ is discussed.
\end{abstract}

\pacs{75.25.+z, 75.30.Gw, 75.40.Cx}

\maketitle


NdRhIn$_5$ crystallizes in the tetragonal HoCoGa$_5$ structure (Space group $P4/mmm$)~\cite{grin1} and belongs to a large structural family of compounds with the chemical composition $R_mM$In$_{3m+2}$, with $R=$~rare earth, $M=$~transition metal and $m=1,2$. The tetragonal crystal structures of these compounds may be seen as $m$ layers of $R$In$_3$ and a layer of $M$In$_2$ alternately stacked along the $c$~axis. Included in this family are three newly discovered heavy-Fermion superconductors that have received considerable attention.~\cite{hegger1,petrovic1,petrovic2} For example, CeRhIn$_5$, an antiferromagnet below $T_N = 3.8$~K, undergoes a transition to a superconducting state at approximately 16~kbar with $T_C = 2.1$~K.~\cite{hegger1,kawasaki1} Another member, CeCoIn$_5$ is an ambient pressure superconductor with a record setting $T_C = 2.3$~K for heavy-Fermion superconductors.~\cite{petrovic2} Thermodynamic and transport measurements are indicative of unconventional superconductivity in which there are line-nodes in the superconducting gap.~\cite{movshovich1,izawa1}

It is widely held that magnetic ground states of heavy-Fermion compounds are determined by the balance between competing Kondo and RKKY interactions.~\cite{doniach1} For \emph{f}-electron magnetic materials, anisotropy is also known to affect the magnetic state.~\cite{jensen1} Therefore, studies of structurally related non-Kondo magnetic materials such as NdRhIn$_5$ may give insight into the evolution of magnetic properties in these materials.~\cite{pagliuso1} For the present study, we have performed both powder and single crystal neutron diffraction in order to determine the magnetic structure of the antiferromagnet NdRhIn$_5$. The results are compared to those of cubic NdIn$_3$, which may be considered the parent compound in the Nd$_mM$In$_{3m+2}$ series. Further comparisons are also made with the evolution of magnetic structures in the Ce-based series.

Single crystals of NdRhIn$_5$ were grown from an In flux. The lattice parameters are $a = 4.630$~\AA~and $c = 7.502$~\AA~at room temperature.~\cite{pagliuso1} Neutron diffraction experiments were performed at Chalk River Laboratories using the C-2 High Resolution Powder Diffractometer and the C-5 triple axis spectromer in a two axis mode.  Incident neutrons of wavelength 1.33~\AA~were selected using a Si monochromator for C-2, while 1.53~\AA~neutrons were selected with a Ge monochromator for C-5. In both cases, the sample temperature was regulated by a top loading pumped He cryostat.

In order to determine the magnetic propagation vector, powder diffraction patterns were collected above and below the ordering temperature using C-2. The low temperature pattern clearly shows additional magnetic reflections which can be indexed using a magnetic structure with the propagation vector $\mathbf q_M =
(\frac{1}{2}~0~\frac{1}{2})$. This corresponds to a magnetic unit cell that doubles the chemical unit cell along the tetragonal $a$ and $c$~axes and contains four magnetic Nd ions.

Subsequently, a rectangular plate-like sample of dimensions $\sim\!3 \times 3 \times 0.7$~mm$^3$ with the $(001)$~plane the largest surface was measured on C-5.  The sample was mounted with the $[010]$ direction vertical in order to access reciprocal lattice points of the type $(h0l)$.

\begin{figure}[!tb]
\includegraphics[width=8.6cm]{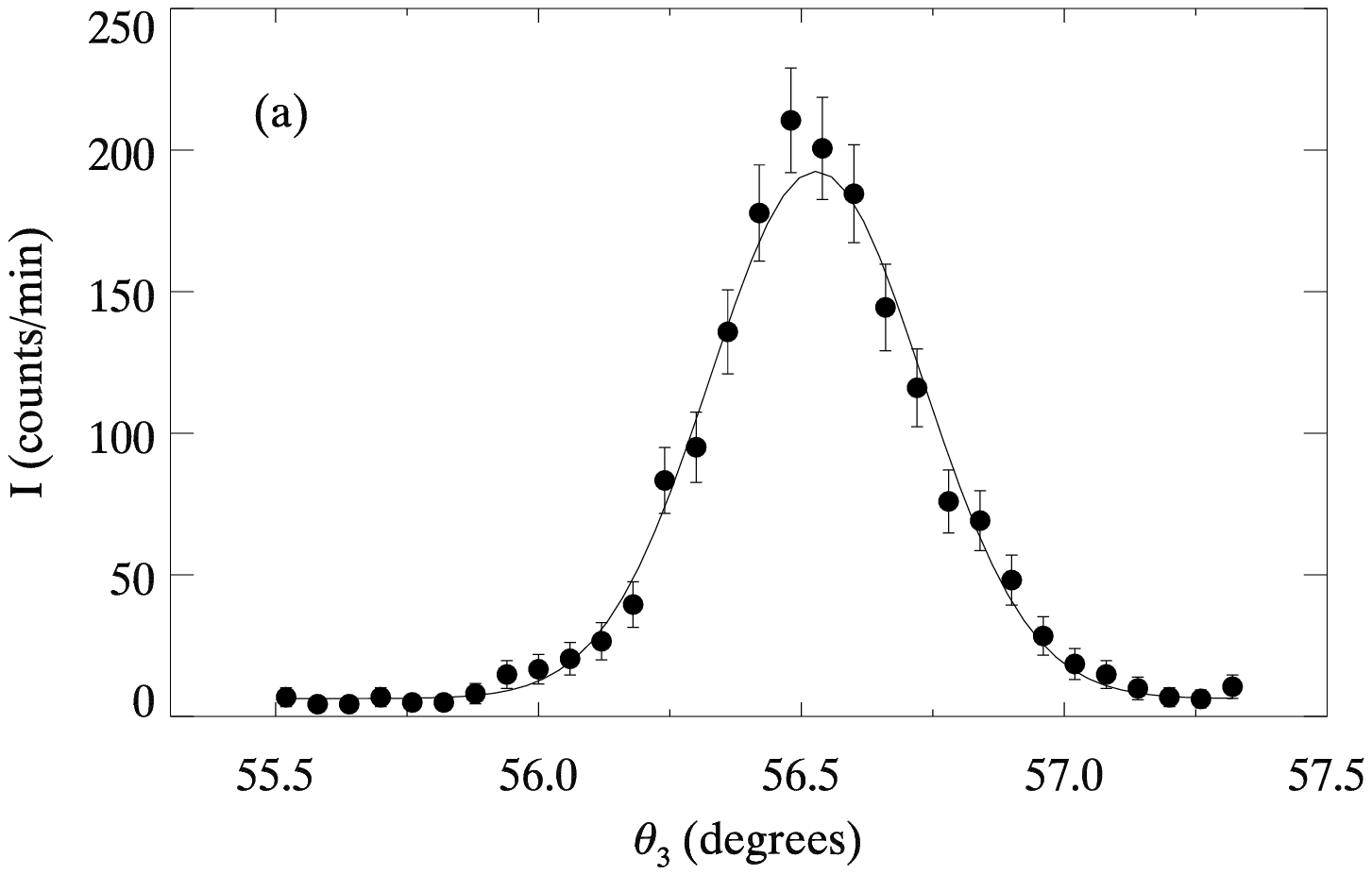}\\
\includegraphics[width=8.6cm]{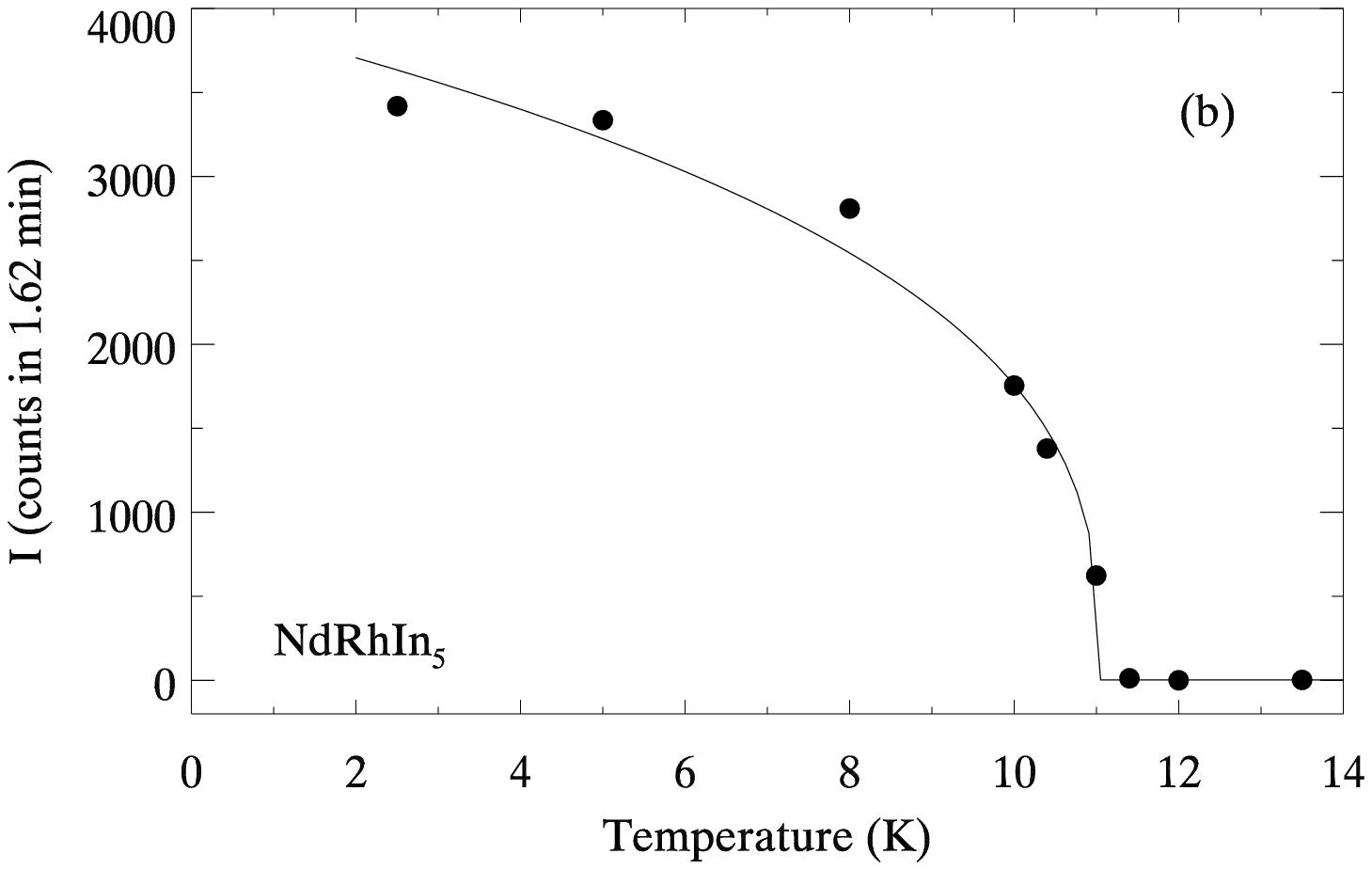}
\caption{(a) Elastic rocking scan through magnetic Bragg point $(\frac{1}{2}~0~\frac{5}{2})$ at 1.6~K. (b) Intensity of $(\frac{3}{2}~0~\frac{1}{2})$ reflection as a fuction of temperature.  The N\'eel temperature is 11~K. The solid line is a guide for the eye}
\label{rocking scan}
\end{figure}

We observed temperature dependent magnetic Bragg reflections at ($\frac{m}{2}~0~\frac{n}{2}$), where $m$ and $n$ are odd integers, confirming the propagation vector found in powder diffraction. A typical elastic rocking scan taken at 1.6~K is shown in Fig.~\ref{rocking scan}(a). The intensity of the $(\frac{3}{2}~0~\frac{1}{2})$ peak is shown in Fig.~\ref{rocking scan}(b) as the square of the order parameter of the antiferromagnetic transition. The N\'eel temperature was determined to be 11.0(1)~K, in good agreement with $T_N$ found inspecific heat measurements.~\cite{pagliuso1} The integrated intensities of magnetic Bragg reflections from such rocking scans were normalized to the (400) and (004) nuclear peaks to yield magnetic cross sections $\sigma_{obs}(\mathbf q) = I(\mathbf q) \sin (2\theta)$ in absolute units. The propagation vector $\mathbf q_M$ suggests a model in which Nd moments are aligned antiferromagnetically in the $[100]$ and $[001]$ directions, and ferromagnetically in the $[010]$ direction, resulting in the magnetic cross section~\cite{squires}
\begin{equation} \label{eq:sigma1}
\sigma(\mathbf q) = \left(\frac{\gamma r_0}{2}\right)^2 \langle m \rangle^2 \vert \mathit f(q) \vert^2 \langle 1 - (\hat{\mathbf q} \cdot \hat{\mathbf m})^2 \rangle,
\end{equation}
where $\gamma r_0/2 = 0.2695 \times 10^{-12}~\mathrm{cm}/\mu_B$ is the scattering length associated with $1~\mu_B$, $\langle m \rangle$ is the staggered moment of the Nd ion, and $\mathit f(q)$ is the Nd$^{3+}$ magnetic form factor.~\cite{blume1} The polarization factor $\langle 1 - (\hat{\mathbf q} \cdot \hat{\mathbf m})^2\rangle$, averaged over possible magnetic domains with the assumption of equal occupation of the domains is
\begin{equation} \label{eq:polarization}
\langle 1 - (\hat{\mathbf q} \cdot \hat{\mathbf m})^2 \rangle = 1 - \frac{\sin^2\alpha \sin^2 \beta + 2 \cos^2 \alpha \cos^2\beta}{2},
\end{equation}
where $\alpha$ is the angle between $\mathbf q$ and the $c$ axis, and  $\beta$ is the angles between the magnetic moment and the $c$ axis. The best least squares fit to Eqs.~(\ref{eq:sigma1}) and (\ref{eq:polarization}) gives, within one standard deviation, $\beta = 0$, which corresponds to magnetic moments aligned along the \textit c~axis, and reduces Eq.~(\ref{eq:sigma1}) to
\begin{equation} \label{eq:sigma2}
\sigma (\mathbf q) = \left( \frac{\gamma r_0}{2} \right)^2 \langle m \rangle^2 \vert \mathit f(q) \vert^2 (1 - \cos^2\alpha).
\end{equation}

\begin{figure}[!tb]
\includegraphics[width=8.6cm]{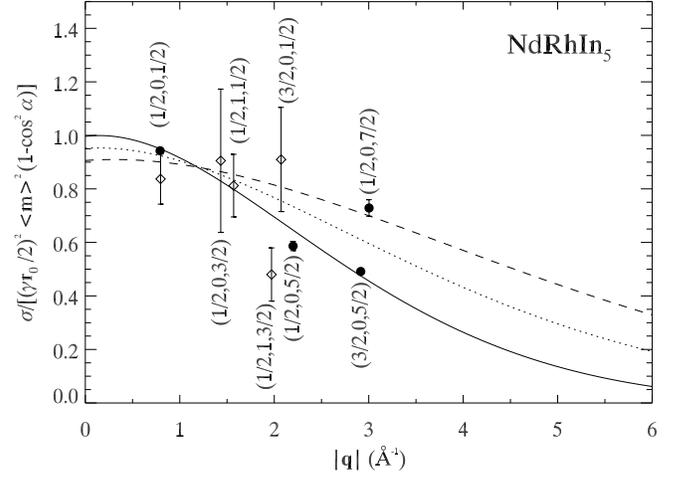}
\caption{The $q$ dependence of the square of the magetic formfactor, $\vert f \vert^2$  as given by the quantity, $\sigma(q)/[(\gamma r_0/2)^2 \langle m \rangle^2 (1 - \cos^2\alpha)]$ [refer to Eq.~(\ref{eq:sigma2})]. The value of the magnetic moment was taken from the best least squares fit of the data. Single-crystal data are shown as solid circles, while data from the powder diffraction experiment are represented by open diamonds. The lines represent the spin-only (solid), orbit-only (dot) and spin+orbit (dash) Nd$^{3+}$ form factors from Ref.~\onlinecite{blume1}.}
\label{formfactor}
\end{figure}

The best least squares fit of the experimental data was achieved using the spin-only Nd$^{3+}$ form factor.~\cite{blume1} The staggered Nd moment is determined at 1.6~K to be $m = 2.61(1)~\mu_B$. Fig.~\ref{formfactor} shows the quantity $\sigma(q)/[(\gamma r_0/2)^2 \langle m \rangle^2(1-\cos^2 \alpha)]$, which is equal to the square of the magnetic form factor, $\vert f \vert^2$ [refer to Eq.~(\ref{eq:sigma2})]. The high temperature effective moment of 3.66~$\mu_B$, deduced from susceptibility data is in good agreement with the Hund's rule value of 3.62~$\mu_B$, indicating well localized Nd moments.~\cite{pagliuso1} One may thus expect a significant orbital contribution to the magnetic moment, although this contribution is likely to be complicated by the presence of CEF effects and related magnetic anisotropy. If an admixture of orbital moments is allowed, the staggered moment can be reduced by ~10\%, as indicated by the intercepts of the dotted and dashed lines in Fig.~\ref{formfactor}, which represent orbit-only and spin+orbit formfactors,~\cite{blume1} taking into consideration the fact that the data points reflect Nd moments determined from the best least squares fit, \emph{i.e.}, the spin-only formfactor.

Cubic NdIn$_3$ orders antiferromagnetically below $T_N = 6$~K and exhibits a complex magnetic phase diagram, which includes two additional antiferromagnetic transitions at 4.61~K and 5.13~K.~\cite{buschow1,czopnik1,czopnik2} The two intermediate phases were determined to have incommensurate structures with magnetic propagation vectors $\mathbf q_M = (\frac{1}{2}~0.037~\frac{1}{2})$ and $(\frac{1}{2}~0.017~\frac{1}{2})$, respectively, while the ground state structure was determined to be commensurate with $\mathbf q_M = (\frac{1}{2}~0~\frac{1}{2})$ and staggered Nd moments of approximately $2.0~\mu_B$ with [010] the easy magnetization direction.~\cite{mitsuda1,amara1} The complexity of the magnetic phase diagram of NdIn$_3$ is also verified by field induced transitions in the H-T phase diagram.~\cite{mitsuda1,amara1} A model including competing CEF and magnetic exchange anisotropies have satisfactorily described the complex phase diagrams of NdIn$_3$.~\cite{mitsuda1,amara1}

\begin{figure}[!tb]
\includegraphics[width=8.6cm]{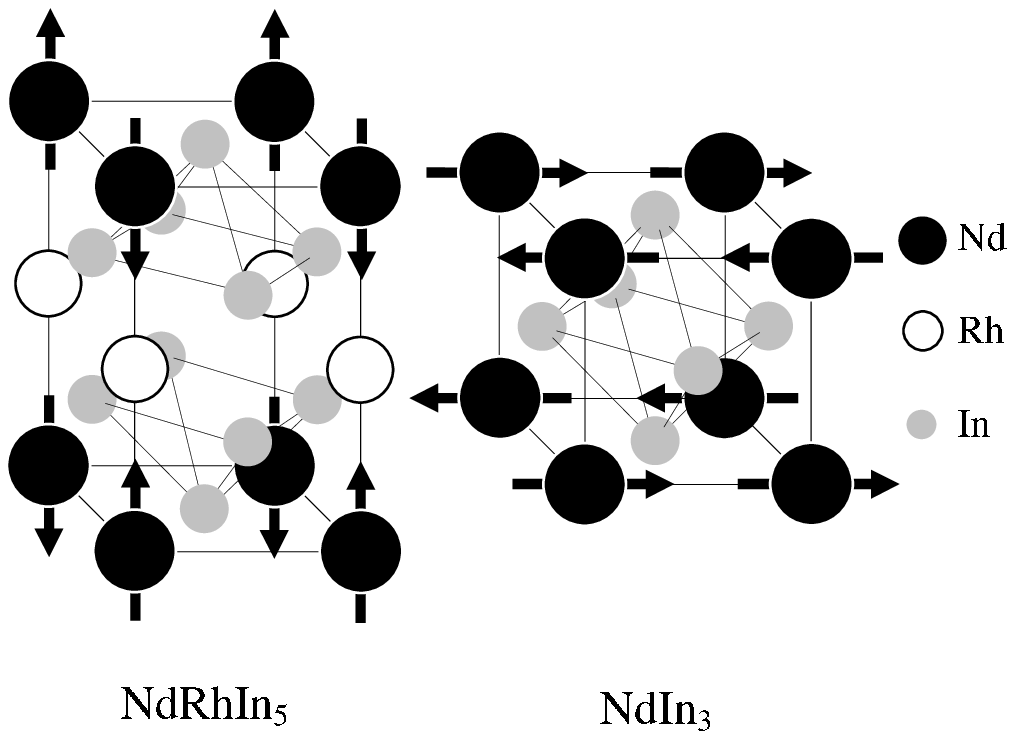}
\caption{Schematic representation of the crystallographic and magnetic structure of NdRhIn$_5$ in a chemical unit cell. The commensurate magnetic structure of NdIn$_3$ below 4.6~K from Ref.~\onlinecite{amara1} is also shown for comparison. The arrows indicate the directions of the Nd moments.}
\label{structure}
\end{figure}

The magnetic structure of NdRhIn$_5$ is shown together with that of  NdIn$_3$ below 4.6~K in Fig.~\ref{structure}. In comparison to NdIn$_3$, the moment direction relative to the magnetic wave vector is rotated by $90^\circ$ in NdRhIn$_5$. However, the phases among the magnetic moments are identical in both cases.

For the tetragonal NdRhIn$_5$, the insertion of a RhIn$_2$ layer nearly doubles the N\'eel temperature of NdIn$_3$. No evidence of additional transitions below $T_N$ was observed in our study as well as in bulk measurements down to 1~K.~\cite{pagliuso1} In addition, field dependent heat capacity revealed no evidence for field induced transitions up to $H = 9$~T applied in the $ab$ plane and one transition at about 7 T for $H \parallel c$~axis.~\cite{pagliuso3} Therefore, although the magnetic structure of NdRhIn$_5$ is closely related to the parent compound NdIn$_3$, the relatively simple H-T phase diagram of NdRhIn$_5$ suggesta that the commensurate antiferromagnetic structure $\mathbf q_M = (\frac{1}{2}~0~\frac{1}{2})$ is more robust and stable in the tetragonal variant. In fact, the Nd$^{3+}$(J=9/2) ion in axial symmetry commonly has its multiplet split in anisotropic doublets (with $g$-value $g_{\parallel c}$ $\gg$ $g_{\perp}$) favoring the Nd spins to point along the $c$~axis which is consistent with our results. Therefore, the tetragonal symmetry may produce an improved matching among the existing CEF, magneto-crystalline and exchange coupling anisotropies for NdRhIn$_5$.

We now extend our discussion to the Ce-based series. For CeRhIn$_5$, the nearest neighbor antiferromagnetic structure of the parent compound CeIn$_3$ is maintained within the CeIn$_3$ layers. However, the magnetic moments in CeRhIn$_5$ form an incommensurate spiral along the $c$~axis.~\cite{bao1,curro1} Furthermore, $T_N$ is reduced by a factor of 2 for CeRhIn$_5$ ($T_N \sim 4$~K) compared to CeIn$_3$ ($T_N \sim 10$~K), which is just the opposite of the situation in NdRhIn$_5$ and NdIn$_3$.

These contrary behaviours can be understood by noticing that the magnetic moments in CeRhIn$_5$ lie in the $ab$ plane,\cite{bao1} whereas the CEF anisotropy tends to favor energetically the Ce spins to point along the $c$~axis.\cite{hegger1,pagliusocef} Therefore, there might be in CeRhIn$_5$ competing anisotropic magnetic interactions that lead to an incommensurate  magnetic state at lower $T_N$ when compared to CeIn$_3$. Accordingly, field dependent heat capacity\cite{Andy} has revealed a rich H-T phase diagram with field-induced transitions similar to what was observed in NdIn$_3$, where competing CEF and exchange interaction anisotropies were considered.

Alternatively, antiferromagnetic correlations across the intervening RhIn$_2$ layers in NdRhIn$_5$ are in some sense more reminiscent of Ce$_2$RhIn$_8$, in which the magnetic structure within CeIn$_3$ bilayers are unmodified relative to cubic CeIn$_3$ and the correlations across the RhIn$_2$ layers are antiferromagnetic.~\cite{bao3}

In conclusion, we find a commensurate antiferromagnetic structure, as represented in Fig.~\ref{structure}, with $\mathbf q_M = (\frac{1}{2}~0~\frac{1}{2})$ for NdRhIn$_5$.  The staggered Nd moment is determined at 1.6~K to be 2.6~$\mu_B$ aligned along the tetragonal $c$~axis. The phases of the nearest neighbor Nd atoms are the same as in the commensurate phase of cubic NdIn$_3$, indicating the strong influence of the NdIn$_3$ building blocks on the magnetic correlations. The enhanced $T_N$ for NdRhIn$_5$ is interpreted in terms of an improved matching among the existing magnetic anisotropies. The same effect in the opposite direction is speculated to play a role in the magnetic properties of CeRhIn$_5$ when compared to CeIn$_3$.

\begin{acknowledgments}
Work at Los Alamos was performed under the auspices of the US Department of Energy.
\end{acknowledgments}

\bibliography{references}

\end{document}